\input miniltx
\input graphicx.sty

\magnification=1200
\baselineskip=13pt
\vsize=9.4truein
\hsize=6.0truein
\overfullrule0pt
\hoffset=0.25in
\voffset=0.0in
\overfullrule0pt
\parskip=4pt
\parindent=20pt
\nopagenumbers
\pageno=1
\font\ninerm=cmr9
\headline={\ifnum\pageno>1\hfill\ninerm\folio\else\fi}

\def\nh{\noindent\hangindent=1.5truecm\hangafter=1}

\def\bs{\bigskip}
\def\cl{\centerline}
\def\ms{\medskip}
\def\ni{\noindent}
\def\ve{\vfill\eject}
\def\vend{\ve\end}

\def\A{{\rm A}}
\def\a{\alpha}
\def\argmin{{\rm argmin}}

\def\be{\beta}
\def\bep{{\bar\ep}}

\def\bX{{\bar X}}
\def\bxi{{\bar\xi}}
\def\bY{{\bar Y}}

\def\cH{{\cal H}}
\def\cI{{\cal I}}
\def\cX{{\cal X}}
\def\ca{{\check a}}
\def\cb{{\check b}}
\def\cbe{{\check\be}}

\def\cov{{\rm cov}}
\def\de{\delta}

\def\ep{\epsilon}
\def\etal{{et al.~}}

\def\ha{{\hat a}}

\def\hb{{\hat b}}
\def\hbe{{\hat\be}}
\def\half{^{1/2}}
\def\hJ{{\hat J}}
\def\hK{{\widehat K}}

\def\hom{{\widehat\om}}
\def\hphi{{\hat\phi}}

\def\hsi{{\hat\si}}
\def\hth{{\hat\th}}
\def\hy{{\hat y}}
\def\ij{_{ij}}
\def\inti{\int_\cI}
\def\ka{\kappa}
\def\la{\lambda}
\def\lhs{{\rm LHS}}
\def\mhf{^{-1/2}}
\def\mi{{\,|\,}}
\def\mo{^{-1}}
\def\mod{_{{\rm mod}}}
\def\om{\omega}
\def\oon{{1\over n}}
\def\part{\partial}
\def\ra{\to}
\def\rai{\to\infty}
\def\si{\sigma}
\def\simp{_{{\rm simp}}}
\def\sumi{\sum_i\,}
\def\sumion{\sum_{i=1}^n\,}
\def\sumjoi{\sum_{j=1}^\infty\,}

\def\sumjom{\sum_{j=1}^m\,}
\def\ta{{\tilde a}}
\def\tb{{\tilde b}}

\def\th{\theta}
\def\thf{{\textstyle{1\over2}}}

\def\BIC{{\rm BIC}}

\baselineskip=16.7pt

\cl{\bf TRUNCATED LINEAR MODELS}
\cl{\bf FOR FUNCTIONAL DATA}

\bs

\cl{Peter Hall{\hskip 2.5cm} Giles Hooker {\hskip 0.5cm} }

\cl{$\;\,$University of Melbourne \quad Cornell University}

\bs\bs

\baselineskip=13pt

\ni{\bf Abstract.}
A conventional linear model for functional data involves expressing a response variable $Y$ in terms of the explanatory function $X(t)$, via the model: $Y=a+\int_\cI b(t)\,X(t)\,dt+\hbox{error}$, where $a$ is a scalar, $b$ is an unknown function and $\cI=[0,\a]$ is a compact interval.  However, in some problems the support of $b$ or $X$, $\cI_1$ say, is a proper and unknown subset of $\cI$, and is a quantity of particular practical interest.  In this paper, motivated by a real-data example involving particulate emissions, we develop methods for estimating~$\cI_1$.  We give particular emphasis to the case $\cI_1=[0,\th]$, where $\th\in(0,\a]$, and suggest two methods for estimating $a$, $b$ and $\th$ jointly; we introduce techniques for selecting tuning parameters; and we explore properties of our methodology using both simulation and the real-data example mentioned above.  Additionally, we derive theoretical properties of the methodology, and discuss implications of the theory.  Our theoretical arguments give particular emphasis to the problem of identifiability.

\ms

\ni{\bf Key words and phrases.}  Consistency, functionally equivalent models, identifiability, mean squared prediction error, regression, slope function, statistical smoothing, support interval.

\ni{\bf Short title.}  Truncated functional regression.

\bs\bs

\baselineskip=16.7pt

\cl{\bf 1. INTRODUCTION}

\ni{\sl 1.1.  Linear model for functional data.}
The linear model for functional regression has the form
$$
Y=a+\inti b\,X+\ep\,,\eqno(1.1)
$$
where independent observations of the pairs $(X,Y)$ are made, $X$ is a random function recorded on the interval $\cI$, $a$ and $Y$ are scalars, $b$ is a function defined on $\cI$, and $\ep$ denotes an experimental error with zero mean. In the case of a truncated linear model there are practical reasons to believe that $Y$ depends on $X$ only through the values taken by $X$ on a subinterval $\cI_1=[u,v]$, say, of $\cI$.

Therefore, in place of (1.1), we ask that
$$
Y=a+\int_u^vb\,X+\ep\,.\eqno(1.2)
$$
If $u$ and $v$ in (1.2) are included, along with $a$ and $b$, among the unknowns in the model at (1.2), then the model is no longer linear.  It is, of course, an example of functional linear regression, which we consider immediately below.

\ni{\sl 1.2.  General functional regression.}
In regression we observe independent replicates of the data pair $(X,Y)$, and the relationship between $X$ and $Y$ is modelled~as
$$
Y=g(X)+\ep\,.\eqno(1.3)
$$
Here $g$ is a real-valued function, or a functional if $X$ is a function, and the experimental error $\ep$ satisfies
$$
E(\ep\mi X)=0\,.\eqno(1.4)
$$

Suppose we can parametrise $g$, either in a conventional sense where only a finite number of parameters are involved, or in a nonparametric setting where the number of parameters is countably infinite. In the first of these contexts we often estimate unknown parameters by minimising an empirical version of the mean squared prediction error,
$$
D_1(g\mod)=E\{Y-g\mod(X)\}^2\,,\eqno(1.5)
$$
where $g\mod$ represents a model that, in cases where $X$ is a random function, might be particularly complex. In nonparametric settings we typically do the same, after disregarding all but $m$, say, of the unknown parameters, and letting $m$ increase with sample size.

The attraction of minimising $D_1(h)$ is that, under a condition such as (1.4), $D_1(g\mod)$ equals the mean squared difference between the true $g(X)$ and the model $g\mod(X)$, the latter expressed as function of unknown parameters, plus the quantity $E(\ep^2)$, which does not depend on the model. Of course, this result does not require the full force of (1.4); it needs only the property that $g(X)-g\mod(X)$ and $\ep$ are uncorrelated, which follows from~(1.4).

The fact that $g\mod$ can be particularly complex motivates consideration of simpler functions, or functionals, alternative to both $g\mod$ and~$g$. These alternatives might be far too simple to capture the true $g$ in any detail, but they can be much simpler to analyse, and hence also simpler to use for prediction.  Importantly, and as we shall show in section~2, these alternative functions include the truncated linear model at (1.2).  This property leads to simple results about the identifiability of that model; see section~2.2.

\ni{\sl 1.3.  Literature survey.}
Methodology for the functional linear model was discussed in Chapter~10 of Ramsay and Silverman (2002), and Chapter~12 of Ramsay and Silverman (2005). Cardot \etal(1999) made a particularly early contribution to the field. Cardot \etal(2003), and Zhang and Chen (2007), discussed the impact of smoothing on inference in the functional linear model; Crambes \etal(2008, 2009), and Maronna and Yohai (2013), introduced methods based on smoothing splines; Ba\'\i llo (2007) suggested kernel techniques and made comparisons with parametric approaches; James \etal(2009) developed variable selection ideas; Mas and Pumo proposed an alternative formulation of the functional linear model; He \etal(2010) introduced techniques based on canonical analysis; Yuan and Cai (2010) suggested a method founded on reproducing kernel Hilbert space analysis; Ferraty \etal(2012) discussed presmoothing methods; and Comte and Johannes (2012), Johannes and Schenk (2012) and Cai and Zhou (2013) treated methods for adaptive smoothing in functional linear regression. Fan and Zhang (2000), Fang \etal(2005) and Wu \etal(2010), among others, developed methodology for functional linear regression in the context of longitudinal data analysis; Cai and Hall (2006) and Apanasovich and Goldstein (2008) addressed mean squared prediction error in functional linear regression; and Cardot \etal (2007), Hall and Horowitz (2007), Li and Hsing (2007), Cai and Yuan (2012) and Johannes and Schenk (2013) discussed convergence rates of estimators of $a$ and $b$ in~(1.1).

\ni{\sl 1.4.  Summary.}
We begin in section~2 by exploring, in general cases including those where $X$ is a function, the class of all candidates for the regression mean $g$ in~(1.3).  In particular, in the setting of the truncated linear model at (1.2), we show that the intercept $a$ and slope function $b$ are identifiable in particularly general circumstances.  This general perspective underpins our development, in section~3, of methodology for estimating $a$, $b$ and the support interval $\cI_1=[u,v]$.  We suggest two methodologies, introduced in parts A and~B, respectively, of section~3.1; in section~3.2 we illustrate the application of those techniques, depending as they do on tuning parameters; and in section~3.3 we introduce methods for choosing the tuning parameters.  Sections~4, 5 and 6 illustrate properties of our methodology through simulation analysis, by application to real data, and through theoretical development, respectively.  Technical arguments are deferred to appendix~A.

\bs

\cl{\bf 2. GENERAL REGRESSION MODELS}

\ni{\sl 2.1. General regression and correlation.}
Let the regression mean $g$ be as at (1.3), and let the alternatives to $g$ be members, $h$ say, of a class~$\cH$. They are appropriate regression models, even when they are incorrect (that is, even when $g\notin\cH$), provided that the version of (1.5) when $g\mod$ is replaced by $h$ can be written as $E\{g(X)-h(X)\}^2$, plus a quantity that does not depend on~$h$. This property, if it were to hold, would reflect the lack of correlation between $g(X)$ and $\ep$ discussed in the previous paragraph, and it is captured by the following:
$$
\eqalign{
&\hbox to4.42in{the version of the experimental error $\ep$ that is implicit in the form of}\cr
\noalign{\vskip-7pt}
&\hbox to4.42in{(1.3) for the new regression problem, is uncorrelated with the fitted}\cr
\noalign{\vskip-7pt}
&\hbox{mean~$h(X)$.}\cr}\eqno(2.1)
$$
We claim that the following constraint is sufficient for~(2.1):
$$
\eqalign{
&\hbox to4.42in{for each $h\in\cH$, and all constants $c_1\in(-\infty,\infty)$ and $c_2>0$, the function}\cr
\noalign{\vskip-7pt}
&\hbox{$c_1+c_2\,h$ is also in~$\cH$.}\cr}\eqno(2.2)
$$
Condition (2.2) is equivalent to asking that each $h$ can be rescaled and recentred at will, without leaving $\cH$. Of course, we require that the quantities $g(X)$, $h(X)$ and $\ep$ have finite variance:
$$
\hbox{$E\{g(X)^2\}+E(\ep^2)<\infty$ and, for all $h\in\cH$, $E\{h(X)^2\}<\infty\,.$}\eqno(2.3)
$$

\proclaim Theorem~2.1. If $(1.4)$, $(2.2)$ and $(2.3)$ hold, then any function $h=h_0\in\cH$ that minimises
$$
D_2(h)=E\{Y-h(X)\}^2\eqno(2.4)
$$
satisfies
$$
E\{g(X)\}=E\{h_0(X)\}\,,\quad
E\big\{h_0(X)^2\big\}=E\{g(X)\,h_0(X)\}\,.\eqno(2.5)
$$

To appreciate why (2.1) follows from Theorem~2.1, note that if we choose $h_0$ to minimise $D_2(h)$ at (2.4), and treat $h_0(X)$ as the new version of $g(X)$, then the model error alters from $\ep$ to $\ep_0=\ep+g(X)-h_0(X)$, and in this setting we can write (1.3) equivalently as $Y=h_0(X)+\ep_0$. Property (2.1) asks that the new $g(X)$, i.e.~$h_0(X)$, and the new error, i.e.~$\ep_0$, be uncorrelated, and it follows from (2.5) that this is indeed the case.

Of course, if $X$ is a random function then (2.2) holds if $\cH$ represents the functional linear model. This is a major attraction of that model---not only is it relatively simple to analyse, but it remains valid as a regression model since its associated noise is uncorrelated with the signal, even if $h$ is not identical to the more intricate functional~$g$.




\ni{\sl 2.2.  Identifiability of the truncated linear model.}
Recall the definition, at (1.2), of the truncated linear model for functional data. In the present section we show that, under a mild condition on the distribution of $X$ (see (2.8) below), a truncated linear model can be identified from data. Consider the possibility that there exists an alternative, functionally equivalent linear model, where the intercept and slope function $a$ and $b$ are replaced by $a_1$ and $b_1$, respectively, and the interval $[u,v]$ is instead~$[u_1,v_1]$:
$$
P\bigg(a+\int_u^v b\,X
=a_1+\int_{u_1}^{v_1}b_1\,X\bigg)=1\,,\eqno(2.6)
$$
where $u,v,u_1,v_1\in\cI$, $u<v$ and $u_1<v_1$.

Suppose too that $\inti E(X^2)<\infty$, and let
$$
K(t_1,t_2)=\cov\{X(t_1),X(t_2)\}=\sumjoi\om_j\,\phi_j(t_1)\,\phi_j(t_2)\eqno(2.7)
$$
denote the singular-value decomposition of the covariance function $K$, where $\om_1\geq\om_2\geq\ldots$ are eigenvalues, and $\phi_1,\phi_2,\ldots$ are the associated eigenfunctions, of the linear operator with kernel~$K$. We assume that:
$$
\eqalign{
&\hbox to4.42in{the linear operator with kernel $K$ is of full rank in $L_2(\cI)$, in the sense}\cr
\noalign{\vskip-7pt}
&\hbox to4.42in{that each $\om_j\neq0$ and the sequence $\phi_1,\phi_2,\ldots$ is complete in the class of}\cr
\noalign{\vskip-7pt}
&\hbox{square-integrable functions on~$\cI$.}\cr}\eqno(2.8)
$$

\proclaim Theorem~2.2. If $(2.6)$ and $(2.8)$ hold then $a=a_1$ and
$$
b(t)\,I(t\in[u,v])
=b_1(t)\,I(t\in[u_1,v_1])\eqno(2.9)
$$
for almost all $t\in\cI$.

To appreciate the implications of Theorem~2.2, suppose the function $b$ is strictly positive on $(u,v)$ and vanishes on $\cI\setminus(u,v)$, and take $u_1$ and $v_1$ to be respectively the supremum and infimum of all candidate values $t$, for $u_1$ and $v_1$ respectively, such that $b_1(t)=0$ for almost all $t\leq u_1$, and $b_1(t)=0$ for almost all $t\geq v_1$. Then (2.9) is equivalent to the assertion that $u=u_1$ and $v=v_1$, and $b=b_1$ almost everywhere on $[u,v]$. If $u_1$ and $v_1$ are defined in this way, and if (2.8) holds, then it follows from Theorem~2.2 that the scalars $a$, $u$ and $v$ are respectively equal to $a_1$, $u_1$ and $v_1$, and $b_1=b$ almost everywhere on~$\cI$.

\ni{\sl 2.3.  Illustration.}
We conclude this subsection with an example showing that truncated linear models sometimes are, unexpectedly, approximations to rather than equivalent to models that are linear but depend on values taken by $X$ only on a subset of~$\cI$. In particular, even if $g$ is linear in $X$ on $\cI$, and even if $g$ depends on $X$ only through the restriction of $X$ to a subinterval $\cI_1$, it may not be possible to represent $g$ as $g(X)=a+\int_{\cI_1}b\,X$, for a scalar $a$ and a function~$b$.

Take $\cI=[0,1]$, $u=0$ and $v=\thf$ for simplicity, and assume that $X(t)=X(1-t)^c$ on $(\thf,1]$, where $c>0$ is a constant. Unless $X\geq0$ on $\cI_1=[0,\thf]$, we should take $c$ to be an integer, but no matter what the sign of $c$ we assume that $c\neq1$. Let $b_1$ and $b_2$ be functions defined on $\cI_1$, and~put
$$
g(X)=\int_{\cI_1}\big\{b_1(t)\,X(t)+b_2(t)\,X(t)^c\big\}\,dt
=\inti b_3(t)\,X(t)\,dt\,,
$$
where $b_3=b_1$ on $[0,\thf]$ and $b_3(t)=b_2(1-t)$ for $t\in(\thf,1]$. This formula presents $g$ as a linear model, but one where $g$ depends only on the restriction of $X$ to~$\cI_1$. Nevertheless it is not, in general, possible to write
$$
g(X)=a+\int_{\cI_1}b(t)\,X(t)\,dt\,,
$$
for a scalar $a$ and function~$b$.

\bs

\cl{\bf 3. METHODOLOGY}

\ni{\sl 3.1. Methodology for estimating $a$, $b$ and interval endpoints.}
We suggest methodology in the case $u=0$, which is the practical setting that motivated our work. In that context we write $\th$ for~$v$. It is assumed that we have independent data pairs $(X_i,Y_i)$, for $1\leq i\leq n$, all distributed as~$(X,Y)$.

Let $\psi_1,\psi_2,\ldots$ denote an orthonormal basis for the class of square-integrable functions on~$\cI$. Then we can write
$$
b(t)=\sumjoi\be_j\,\psi_j(t)\,,\quad
X(t)=\sumjoi\xi_j\,\psi_j(t)\,,\quad
\inti b\,X=\sumjoi\be_j\,\xi_j\,,\eqno(3.1)
$$
where $\be_j=\inti b\,\psi_j$, $\xi_j=\inti X\,\psi_j$, and the first two series in (3.1) represent generalised Fourier representations for $b(t)$ and $X(t)$, respectively. If we truncate the third series after $m$ terms then we obtain an approximation to the regression mean:
$$
E(Y\mi X)\approx a+\sumjom\,\be_j\,\xi_j\,.
$$

Inference in the linear functional regression model often is based on this generalised Fourier approximation. In practice the $\psi_j$s are often chosen to be empirical principal component functions, for example the functions $\hphi_j$ defined by the singular-value decomposition of the empirical covariance function:
$$
\hK(t_1,t_2)=\oon\,\sumion\{X_i(t_1)-\bX(t_1)\}\,\{X_i(t_2)-\bX(t_2)\}
=\sumjoi\hom_j\,\hphi_j(t_1)\,\hphi_j(t_2)\,,\eqno(3.2)
$$
where $(\hom_j,\hphi_j)$ are the (eigenvalue, eigenfunction) pairs associated with the linear operator with kernel $\hK$, $\bX=n\mo\,\sumi X_i$, and terms are ordered such that $\hom_1\geq\hom_2\geq\ldots$.

The expansion (3.2) is an empirical version of (2.7), and $\hom_j$ and $\hphi_j$ are, under mild conditions, root-$n$ consistent estimators of $\om_j$ and $\phi_j$, respectively, in~(2.7). (See Hall and Hosseini-Nasab, 2009.) If the random functions $X_i$ are continuous, in the sense that the expected value of the Lebesgue measure set $\{t\in\cI:X_1(t)=X_2(t)\}$ equals~0, then with probability~1 the functions $\hphi_1,\ldots,\hphi_n$ are orthonormal on $\cI$, reflecting the fact that $\phi_1,\phi_2,\ldots$ in (2.7) are orthonormal on~$\cI$. Since $\hom_j=0$ for $j\geq n+1$ then the $\hphi_j$s are not defined explicitly for $j$ in this range.

In the case of the truncated linear functional regression model, at least two approaches are feasible, as follows.

\ni{\sl A. First method: Simultaneous inference.}
Determine estimators $\ha$, $\hbe_j$ and $\hth$ of $a$, $\be_j$ and $\th$ by minimising the sum of squares,
$$
S_1(a,\be_1,\ldots,\be_m,\th\mi m)
=\sumion\bigg[Y_i-a-\int_0^\th\,\bigg\{\sumjom\be_j\,\psi_j(t)\bigg\}\,X_i(t)\,dt\bigg]^2\,.
\eqno(3.3)
$$
Here $m$ can be viewed as a smoothing, or regularisation, parameter for estimating~$b$; taking $m$ too large produces an estimator, $\hb=\sum_{1\leq j\leq m}\,\hbe_j\,\psi_j$, that suffers from excessive variance, while choosing $m$ too small results in unnecessarily large bias. A truncated linear predictor of $Y$, when $X=x$, is given~by
$$
\hy(x)=\ha+\int_0^\hth \hb(t)\,x(t)\,dt\,,
$$
being an estimator of $y(x)=a+\int_{[0,\th]}b\,x$.

Not unexpectedly, however, this approach is inadequate for estimating $\th$, since it does not encourage the choice of an estimator $\hth$ that is noticeably less than the upper endpoint of the interval~$\cI$. To improve performance in this regard we add a penalty term to $S$, obtaining:
$$
S(a,\be_1,\ldots,\be_m,\th\mi m,\la)=S_1(a,\be_1,\ldots,\be_m,\th\mi m)+n\,\la\,\th^2\,,
\eqno(3.4)
$$
where $\la>0$ is another tuning parameter. We have multiplied $\la$ by $n$ in (3.4) since both $S$ and $S_1$ are of order $n$. The multiplier will assist our intuition when we assess the impact of $\la$, particularly in section~6. The use of the penalty $\th^2$ in (3.4) is motivated by a Laplace approximation  employed in the methods for selecting $\la$ in Section 3.3, but any continuous increasing function of $\ta$ could be used.

If $\la$ is too large then minimising $S$ tends to produce a relatively small estimator $\hth$, whereas if $\la$ is too small then we produce results similar to those obtained by minimising $S_1$, rather than~$S$; that is, $\hth$ is too large. Choice of $m$ and $\la$ is discussed in section~3.3.

\ni{\sl B. Second method: Iterative inference.}
Here we suggest estimating $a$ and $b$ first, obtaining $\ca$ and $\cb$, say, constructed using a standard method; and then estimating~$\th$. Approaches that can be used to compute $\ca$ and $\cb$ are discussed by, for example, Ramsay and Silverman (2005, Chapter~12), Hall and Hosseini-Nasab (2006) and Crambes \etal(2008, 2009); see section~1 for a more detailed account of the literature. In the second step for this method we employ again penalised least-squares, but this time we select $\th=\hth$ to minimise
$$
T(\th)=\sumion\bigg\{Y_i-\ca-\int_0^\th\cb(t)\,X_i(t)\,dt\bigg\}^2
+n\,\la\,\th^2\,.\eqno(3.5)
$$
See section~3.3 for choice of~$\la$.

Having computed $\hth$ we can proceed in at least two ways. Most simply, assuming for notational clarity that $\cI=[0,1]$, we can define $\hb$ by truncation and $\ha$ by correcting $\ca$ in the obvious way for location:
$$
\hb(t)=\cases{\cb(t)&if $t\leq\hth$\cr
                   0&if $t>\hth\,,$}\quad
\ha=\ca+\int_0^\hth\hb\,\bX-\int_0^1\cb\,\bX
=\ca-\int_\hth^1\hb\,\bX\,.
$$
Alternatively we can use a standard method (for example, the one that produced $\ca$ and $\cb$ in (3.5)) to compute new estimators of $a$ and $b$, this time in the linear regression model $E(Y\mi X)=a+\int_0^\hth b\,X$.

\ni{\sl 3.2. Examples of standard methods for estimating $a$ and~$b$.}
Given a complete orthonormal sequence $\psi_1,\psi_2,\ldots$, and an integer $m\geq1$, the scalar $a$ and function $b$ typically are defined by minimising
$$
\sumion\bigg[Y_i-a-\inti\bigg\{\sumjom\be_j\,\psi_j(t)\bigg\}\,X_i(t)\,dt\bigg]^2\,;
$$
compare (3.3). This results in $\cb=\sum_{j\leq m}\,\cbe_j\,\psi_j$ and $\ca=\bY-\inti\cb\,\bX$, where $\bX$ is as defined in section~3.1, $\bY=n\mo\,\sumi Y_i$, and $\cbe_1,\ldots,\cbe_m$ solve the linear system of equations
$$
\sumjom\be_j\inti\!\inti\psi_k(t_2)\,\psi_j(t_1)\,\hK(t_1,t_2)\,dt_1\,dt_2
=\inti R(t)\,\psi_k(t)\,dt\eqno(3.6)
$$
for $k=1,\ldots,m$, with
$$
R(t)=\oon\,\sumion(Y_i-\bY)\,\{X_i(t)-\bX(t)\}\,.
$$
Typically $m$ is determined by cross-validation or an information criterion.

If we take $\psi_j=\hphi_j$ for $j=1,\ldots,m$, where $m\leq n$ and $\hphi_j$ is as in (3.2), then in view of the orthonormality of those functions, (3.6) simplifies conveniently to
$$
\be_k=\cbe_k={1\over\hom_k}\inti R(t)\,\hphi_k(t)\,dt\,,
$$
where $\hom_k$ is as in (3.2). Equivalently,
$$
\cbe_k={1\over n\,\hom_k}\,\sumion
\{E(Y_i\mi X_i)+\ep_i\}\inti(X_i-\bX)\,\hphi_k
=B_k+\oon\,\sumion B_{ik}\,\ep_i\,,\eqno(3.7)
$$
where $B_k=(n\,\hom_k)\mo\,\sumi E(Y_i\mi X_i)\inti(X_i-\bX)\,\hphi_k$ and $B_{ik}=\hom_k\mo\,\inti(X_i-\bX)\,\hphi_k$, and where $B_k,B_{k},\ldots,B_{kn}$ are all measurable in the sigma-field, $\cX$ say, generated by $X_1,\ldots,X_n$.

The latter technical property has helpful implications, both practical and theoretical. In regression we undertake inference conditional on the design variables, and so the only source of variability comes from the experimental errors~$\ep_i$. Result (3.7) tells us that the estimated Fourier components depend linearly in the $\ep_i$s, with coefficients depending only on the $X_i$s, and in particular that $E(\cbe_j\mi\cX)=B_j$ and $E(\cb\mi\cX)=\sum_{j\leq m}\,B_j\,\hphi_j$. This ensures a simple, equivalent reformulation of $T(\th)$, at~(3.5):
$$
\eqalignno{
T(\th)&=\sumion\bigg\{E(Y_i\mi X_i)-\ca-\int_0^\th\cb(t)\,X_i(t)\,dt\bigg\}^2\cr
&\qquad
+2\,\sumion\bigg\{E(Y_i\mi X_i)-\ca-\int_0^\th\cb(t)\,X_i(t)\,dt\bigg\}\,(\ep_i-\bep)
+\sumion(\ep_i-\bep)^2
+n\,\la\,\th^2\cr
&=\sumion\bigg\{E(Y_i\mi X_i)-\ca-\int_0^\th\cb(t)\,X_i(t)\,dt\bigg\}^2\cr
&\qquad
+2\,\sumion\bigg\{E(Y_i\mi X_i)
-\sumjom B_j\int_0^\th\hphi_j(t)\,X_i(t)\,dt\bigg\}\,(\ep_i-\bep)\cr
&\qquad
-{2\over n}\,\sumion\ep_i\,(\ep_i-\bep)\,\sumjom B\ij\int_0^\th\phi_j(t)\,X_i(t)\,dt
+\sumion(\ep_i-\bep)^2+n\,\la\,\th^2\,.&(3.8)\cr
}
$$

\ni{\sl 3.3. Algorithms for selecting tuning parameters.}
The tuning parameter $\lambda$ plays an important role regularizing the choice of $\theta$. However, because we expect that $\cb(t)$ will be near zero when $t > \theta$, standard methods for choosing $\lambda$ are unlikely to yield good performance results.

Instead, we propose selecting $\lambda$ based on our ability to reconstruct a parametric model $\tb\simp(t)$ intended to approximate $b(t)$. We begin by computing an approximate mean squared error for reconstructing $\tb\simp(t)$, using each of Method~A and Method~B in Section 3.1 for each $\lambda$. We then choose the $\lambda$ that minimises this error and apply it in the original problem. As we shall show in Section 7, our choice of the parametric form for $\tb\simp(t)$ has little effect on the resulting estimators.

After selecting $\lambda$, the number of orthogonal components, $m$, to be employed when using Method~A is selected by~BIC. In the case of Method~B this selection takes place within the estimate of $\cb(t)$.

To be specific, implementation of our method for Method~A involves the following steps. Method~B differs only in Steps (A4) through~(A6):

\ni(A1) Construct the pilot estimators $\ca$ and $\cb(t)$ without a truncation constraint---these will later be employed in Method~B.

\ni(A2) From the estimators $\ca$ and $\cb(t)$,  compute residuals $\tilde{\epsilon_i} = Y_i - \ca - \int_{\cI} \cb X_i$ and compute an empirical variance $\hat{\sigma}^2 = n\mo\,\sum_{i=1}^n \tilde{\epsilon}_i^2$.

\ni(A3) Compute a parametric estimate $\tb\simp$. This can be a straight line or parametric curve designed ``by eye'' to mimic $\cb(t)$ and to decrease to 0 and strike the $t$ axis at~$\bar{\theta}$, say.

    In our simulation analysis in section~4 we shall use a parametric approximation in terms of a Fourier basis:
    $$
    \tb\simp(t) = \left\{ c_0 + c_1 \sin\big( 2^k \pi t\big) + c_2 \cos\big(2^k \pi t\big) \right\}\, I(t < \theta)\,,
    $$
    where $k=1$, 2 or 3 and we estimate $(c_0,c_1,c_2,\theta)$ by minimising unpenalised squared error; see~(3.3).  We could also choose other low-dimensional representations, such as a polynomial basis.  For a fixed-dimensional representation, it is easy to show that these estimates are asymptotically unbiased if the true $b$ falls within the model class.

\ni(A4) For each $\theta$ we can obtain expressions for the mean squared prediction error and mean squared error for $\hat{b}$ at truncation point $\theta$ when using data from the parametric model,
    $$
        Y_i^* = \ca + \int_{\cI} \tb\simp\,X_i + \epsilon_i^*\,,   \eqno(3.9)
    $$
    where the $\epsilon_i^*$s are distributed as normal N$(0,\hat{\sigma}^2)$. We shall also use the noiseless expected values $\bar{Y}_i^* = \ca + \int_{\cI} \tb\simp \,X_i$.

    Specifically, we employ $\psi_j = \hat{\phi}_j^{\theta}$, the empirical Fourier components for the truncated functions $X_i(t)\,I(t < \theta)$ (i.e.~the $X_i$s restricted to the range $[0,\theta]$) which have associated variance components $\tau_j^{\theta}$. We then obtain estimates $a^*_{\theta}$ and $b^*_{\theta}$ by minimising the squared error for predicting the noiseless data $\bar{Y}_i^*$, and define predicted values $\hat{Y}_i^*(\theta) = a^*_{\theta} + \int_{\cI} b^*_{\theta}\,X_i$.

\ni(A5) For $\hat{a}^*_{\theta}$ and $\hat{b}^*_{\theta}$ estimated from the (hypothetical) data $Y_i^*$, the mean squared prediction error~is
    $$
    S^A_{Y}(\theta) = E\bigg\{\sum_{i=1}^n \left( Y_i^*
    - \hat{a}^*_{\theta} - \int_{\cI} \hat{b}^*_{\theta} X_i \right)^{\!2}\;\bigg|\;\cX\bigg\}
    = \sum_{i=1}^n \left\{ \bar{Y}_i - \hat{Y}_i^*(\theta) \right\}^2 + \hat{\sigma}^2\,(m+1)\,,
    $$
    and the mean squared error for estimating $\tb\simp$ is given by
    $$
    S^A_b(\theta) = E\bigg\{ \int_{\cI} \left( \tb\simp
    - \hat{b}^*_{\theta} \right)^{\!2}\;\bigg|\,\cX\bigg\}
    = \int_{\cI} \left( \tb\simp - b^*_{\theta} \right)^{\!2} + \hat{\sigma}^2\,\sum_{j=1}^m\,\left(\tau_j^{\theta}\right)^{-1}.
    $$

\ni(A6) For every $\lambda$ we select $\theta$ to minimise the expected value of (3.4):
$$
\theta_{\lambda} = \argmin_\th\,\big\{S^A_Y(\theta) + \lambda\,\theta^2\big\}\,.
$$
Further, estimate the variance of this choice by
$$
V(\lambda) = {\hsi^2\over n}\;{(d^2/d\theta^2)\, S^A_Y(\theta_{\lambda}) \over
\left\{(d^2/d \theta^2)\,S^A_Y(\theta_{\lambda}) + \lambda \right\}^2 }\;.
$$

\ni(A7) We now choose $\lambda$ to minimise the expectation of $S_b^A(\theta)$ with respect to a normal distribution for $\theta$ with mean $\theta_{\lambda}$ and variance $V(\lambda)$:
$$
P_{b}(\lambda) = \int_{\cI} S_{b}^A(\theta)\;{{1}\over{\sqrt{2 \pi V(\lambda)}}}\;
\exp\big\{-(\theta - \theta_{\lambda})^2/2 V(\lambda) \big\}\,d \theta\,.
$$
For this $\lambda$ we determine $\hat{a}$ and $\hat{b}$ from the original data via Method~A for each choice of $m$. We repeat the process above for each $m$, and select $m$ by minimising~BIC:
$$
\BIC(m) = \log\bigg\{{{1}\over{n}} \sum_{i=1}^m \left( Y_i - \hat{a} - \int_{\cI} \hat{b}\,X_i \right)^{\!2}\bigg\}
+ (m+1)\,\log(n)\,.
$$

For Method~B, the process is analogous. We select $m$ via BIC when computing estimates $\ca$ and $\cb$ in Step (A1). Steps (A4) through (A6) are replaced by:

\ni(B4) Obtain estimates $\ca^*$ and $\cb^*$ to estimate the $\bar{Y}_i^*$ without truncation using $\psi_j = \hat{\phi}_j^1$---the empirical Fourier components on the full interval with $m$ chosen as in Step (A1), and set the predicted values at $\theta$ to be $\check{Y}^*_i(\theta) = \ca^* + \int_0^{\theta} \cb^* X_i$.

\ni(B5) Obtain the quantities analogous to $S_Y^A(\theta)$ and $S_b^A(\theta)$ by employing Method~B:
    $$
    S^B_{Y}(\theta) = \sum_{i=1}^n \left\{\bar{Y}_i - \check{Y}_i^*(\theta) \right\}^2
    + \hat{\sigma}^2\,\sum_{j=1}^m \left( \tau_j^1\right)^{-1}\,\sum_{i=1}^n \left( \int_0^{\theta} \phi_j^1 X_i  \right)^2
    $$
    and
    $$
    S^B_b(\theta) = \int_{\cI} \left\{ \tb\simp - \cb^*I(t < \theta) \right\}^2 + \hat{\sigma}^2 \sum_{j=1}^m \left(\tau_j^{\theta}\right)^{-1} \int_0^{\theta} \left(\hat{\phi}_j^1\right)^2.
    $$

\ni(B6) For every $\lambda$ we select $\theta$ to be the first minimum of the expected value of~(3.4):
$$
\theta_{\lambda} = {\rm first \ minimum \ of \ } S^B_Y(\theta) + \lambda \theta^2\,,
$$
and estimate the variance of this choice by
$$
V(\lambda) = {\hsi^2\over n}\;{(d^2/d\theta^2)\, S^B_Y(\theta_{\lambda}) \over
\left\{(d^2/d \theta^2)\,S^B_Y(\theta_{\lambda}) + \lambda \right\}^2 }\;.
$$

We suggest using the first minimum of $S^B_Y(\theta)$ because $\cb$ is already estimated by minimising squared error. There tends to be a sharp drop in $S^B_Y(\theta)$ close to the right hand endpoint of the interval, so that $S^B_Y(\theta) + \lambda \theta^2$ is minimised at $\theta = 1$ unless $\lambda$ is very large.

This scheme is intended to mimic simulating from the model at (3.9), either directly or by bootstrapping the $\tilde{\epsilon}_i$s, but it substantially reduces computational effort. Empirically, $P_b(\lambda)$ approximates the mean squared error for estimating $b$ after choosing the truncation level very well.

Note that we have selected $\lambda$ based only on our ability to estimate $\tb\simp$. We could also have included mean squared error for $a$, and approximated the error for estimating $\theta$ within these calculations as well.  Observe too that, while these estimation schemes appear similar, the need for a separate principal components analysis for each $\theta$ in Method~A represents a significant additional computational cost.

While the methods above rely on empirical orthogonal components calculated from the $X_i$s, other finite-dimensional linear representations for $b$ (e.g.~explicitly using a polynomial or trigonometric basis) can be employed  with some changes to the form of the expected mean squared error calculations above. However, employing Method~A we have found that unless the basis is adapted to each $\theta$, the estimate $\hat{b}$ can become numerically unstable as basis functions designed for the interval $[0, 1]$ become close to collinear when restricted to~$[0,\theta]$.

\bs

\cl{\bf 4. SIMULATION STUDIES}

\ni We expect that the performance of our methods will be affected strongly by the way in which $b$ tends to zero---a discontinuous drop to zero should be easier to detect, while smooth convergence will make it harder to localise the value of $\theta$.  To examine the performance of our methods, we conducted a simulation study. In this study we generated covariates  $X_i$, $i = 1,\ldots,100$, via a trigonometric basis expansion on the range [0,1] given by:
$$
\eta_1 = 1, \ \eta_{2k } = \sin\big( 2^ k\pi t\big)\,, \quad \eta_{2k+1} \cos\big( 2^k \pi t\big)\,.
$$
The $X_i$s were generated via linear combinations of the first 25 such functions, with coefficients chosen as independent mean-zero Gaussian random variables with the coefficient for $\eta_k$ having variance $\exp\{ -(k-1)/4 \}$. This produces exponentially-decaying variance components.

Our estimates of $\cb$ were computed using the empirical principal components of the generated $X_i(t)$, and we used two through nine of these, choosing the number by~BIC. In Method~A, we employed a separate principal components analysis for each value of~$\theta$.

The methods also rely on a parametric approximation to $\tb\simp$. For this, we employed the first two components of the trigonometric basis,
 $$
 \tb\simp(t,\beta_1,\beta_2,\theta) = \left\{\beta_1 \psi_1(t) + \beta_2 \psi_2(t)\right\}\,I_{t \leq \theta}\,.
$$

Finally, we considered three simulation settings, each with $\theta = 0.5$---the mid-point of the interval. A true $b$ for each was given by the following parametric models:

\ni{\bf Model 1} $b(t) = \psi_1(t)\,I_{t \leq \theta} = I_{t \leq 0.5}$,

\ni{\bf Model 2} $b(t) = \psi_2(t)\,I_{t \leq \theta} = \sin( 2 \pi t) I_{t < 0.5}$,

\ni{\bf Model 3} $b(t) = \{\psi_3(t) + \psi_1(t)\}\,I_{t \leq \theta} = \{\cos( 2 \pi t ) + 1\}\, I_{t < 0.5}$.

\ni These three functions have discontinuities at $\theta$ in the 0th, 1st or 2nd derivatives, representing important behaviour at the $t = \theta$ boundary.  Each of these was scaled to give a 26.25/1 signal to noise ratio for additive Gaussian noise with unit variance added to the observations.

Note that in this case, for Model 1 and Model 2, the form of $\tb\simp$ includes the model class, but it does not for Model 3. We also experimented with specifying $\tb\simp$ using one and three components of the trigonometric basis system, and found that for all three models, the choice of $\theta$, and hence the estimate of $\hat{b}$, was almost always identical for any choice of $\tb\simp$ on a given data set---although the selection of $\lambda$ varied somewhat.  We have therefore only presented one of these.

We ran 400 simulations for each of the three models using $n = 100$. Table 1 presents the mean and standard deviation of the estimates of $\theta$ for Method~A and Method~B for each model. Here we see that Method~B produces estimates that are less variable than those of Method~A, but exhibits more bias. The expected degradation of our estimates when $b$ tends to zero more smoothly is apparent in the observed bias towards more truncation, rather than in the variance. Employing a tapered estimate may improve this.


\bs

\cl{TABLE 1}
\moveright 1cm
\vbox{\offinterlineskip
\vskip 0.25cm
\halign{\strut\vrule \quad # \hfil & \vrule \quad  # \hfil &  \quad # \hfil & \vrule  \qquad # \hfil &  \quad # \hfil \vrule \cr \noalign{\hrule}
& Method~A &  & Method~B  & \cr \noalign{\hrule}
& Mean & Std. Dev. & Mean & Std. Dev \cr \noalign{\hrule}
Model 1 & 0.5197 & 0.0686 & 0.5094 & 0.0156 \cr
Model 2 & 0.4933 & 0.0823 & 0.4799 & 0.031 \cr
Model 3 & 0.3734 & 0.0649 & 0.3527 & 0.0185 \cr \noalign{\hrule}
}}
{\leftskip = 1cm
\rightskip = 1cm
\ni Mean and standard deviation of estimates of $\theta$ from Method~A and Method~B based on 400 simulations for each of three models in which $b(t) = 0$ for $t > 0.5$.
\par
}


Turning to the estimate of $b$ itself, Table 2 presents mean integrated squared error and median integrated
squared error for both Method~A and Method~B as well as for $\cb$ employing no truncation for all three simulation models. Here we see that employing truncation results in an improvement for both methods, and that Method~A improves on Method~B for Models 1 and 3, largely due to the bias that $\cb$ exhibits when trying to estimate a function that is identically zero on part of its domain.  There is an important distinction in performance for Method~A between mean and median squared error, particularly in Model 3; this is due to BIC occasionally selecting a very large $m$, yielding a high-variance estimate that can be diagnosed readily.



\cl{TABLE 2}
\cl{Mean Squared Error}
\moveright 2cm
\vbox{\offinterlineskip
\vskip 0.25cm
\halign{\strut\vrule  \hskip 0pt # \hfil & \vrule \quad \hfil # & \quad \hfil #  & \quad \hfil #  \vrule  \cr \noalign{\hrule}
        & Method~A    & Method~B  & No Trunc.  \cr \noalign{\hrule}
Model 1 & 572.6798 & 800.2337 & 1153.9327 \cr
Model 2 & 548.0986 & 172.1635 & 317.7989 \cr
Model 3 & 1139.5829 & 1037.504 & 1671.0593  \cr \noalign{\hrule}
}}

\cl{Median Squared Error }
\moveright 2cm
\vbox{\offinterlineskip
\vskip 0.25cm
\halign{\strut\vrule  \hskip 0pt # \hfil & \vrule \quad \hfil # & \quad \hfil #  & \quad \hfil #  \vrule  \cr \noalign{\hrule}
        & Method~A & Method~B & No Trunc. \cr \noalign{\hrule}
Model 1 & 441.7294 & 839.5315 & 1305.7375\cr
Model 2 & 154.0908 & 146.0178 & 275.448\cr
Model 3 & 495.4372 & 1131.5556 & 1696.8166 \cr \noalign{\hrule}
}}
{\leftskip = 1cm
\rightskip = 1cm
\ni Mean integrated squared error (left hand columns) and median integrated squared error (right hand columns) for estimating $b$ for each of three models based on 400 simulations.
\par
}

To illustrate these results more concretely, Figure 1 presents plots of the results of the simulation for Model 2. We have provided histograms of the estimated $\hat{\theta}$ for each method, as well as plots of the estimate $\hat{b}$ along with the untruncated~$\cb$.  We see that Method~B's difficulties are largely associated with bias towards small values of~$\theta$, and that there are occasional ``wild'' estimates from Method~A.  These results suggest that Method~B can be employed as a computationally inexpensive means of deciding whether truncation should be attempted before going to the expense of employing Method~A to provide a new estimate.  Results for Models 1 and 3 (not shown) are similar.


\bs

\cl{FIGURE 1}
\moveright 0.5cm
\vbox{
\includegraphics[height=10cm,angle=270]{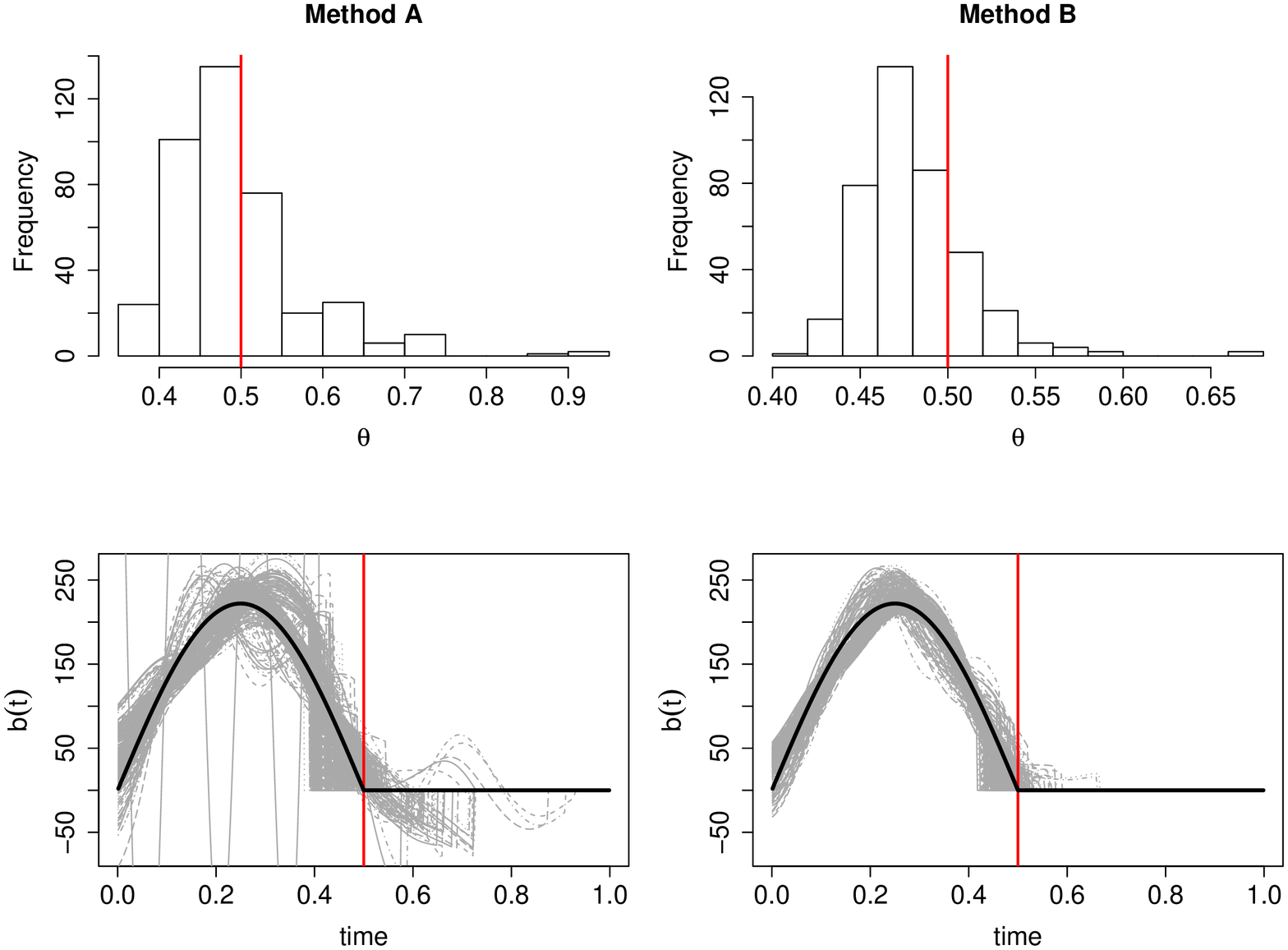}

\includegraphics[height=5cm]{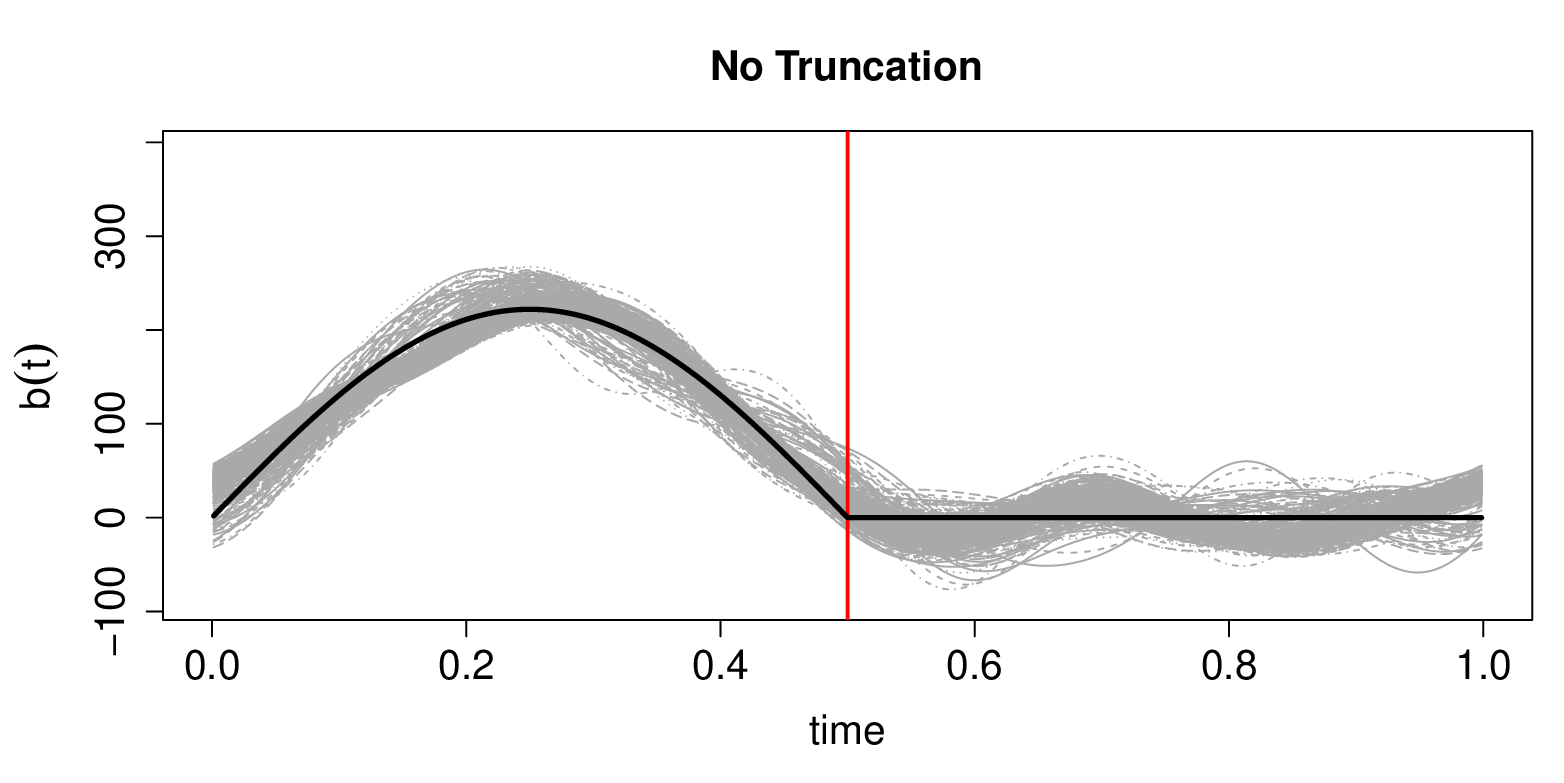}
}
{\leftskip = 0.5cm
\rightskip = 0.5cm
\ni Results of a simulation study with Model 2. Top row: Histograms of $\hat{\theta}$ over 400 simulations; the vertical bar indicates the true values of $\theta = 0.5$. Left column: Method~A; Right column: Method~B.  Second and bottom rows: estimate $\hat{b}$ (grey lines) along with target (dark, thick lines) for Method~A (right), Method~B (left) and without truncation (bottom).  The vertical bar indicates $\theta = 0.5$.
\par
}

\bs

\cl{\bf 5. AN ANALYSIS OF PARTICULATE MATTER EMISSIONS}

The methods developed in this paper are motivated by a problem of modelling particulate matter (PM) emissions from diesel trucks.  For details, see Clark {\it et. al.} (2007). In these data, trucks are placed on stationary rollers and a particle counter is attached to the exhaust pipe of each. The trucks are then driven through a pre-set driving cycle and PM at the tail pipe is measured every second. Asencio {\it et.~al.} (2014) proposed the following model for these data:
$$
\log\{{\rm PM}(t)\} = \int_{0}^{\theta} b(u)\,Z(t-u) \,du + \epsilon(t)\,,
$$
where $Z(t)$ is the acceleration applied by the engine.  That is, $\log$(PM) follows a linear model based on the past $\theta$ seconds of acceleration.  The model is intended to represent mixing of particles in the exhaust pipe.  McLean {\it et. al.} (2014) examined these data for non-linear dependence.

In order to remove dependencies in the data, we have down-sampled PM to obtain an observation every 10 seconds after the first two minutes of data, and have used the previous 60 seconds of acceleration as the corresponding functional covariate. Based on domain knowledge, PM is not expected to take longer than one minute in transport through the exhaust.  That is, we have a data set
$$
Y_i = \log\{{\rm PM}(10 i+120)\}\,, \ X_i(t) = Z(10i + 120 - t)
$$
(note that ``time'' for the stochastic process $X_i$ is now reversed relative to that for~$Z$), where we have 107 observations. The covariates $X_i$ are obtained by smoothing measured velocities in each time window and obtaining a derivative.

Below we illustrate the result of using both Methods~A and~B to obtain estimates of $b$ and $\theta$. We also perform a residual bootstrap based on the results of Method~A, which we used to estimate pointwise standard deviations for each of our estimates and which are represented in the confidence intervals in Figure 2. Here we see that without truncation, the estimate appears to be zero after about 20 seconds. Method~B suggests truncating at 18 to 20 seconds, while Method~A suggests a possibly longer window although the point estimate for truncation is at 13 seconds. The roughness of the confidence intervals in Method~A are associated with the choice of the number of orthogonal components (re-obtained for each bootstrap) which we chose to be between 2 and 9 via~BIC. These results compare with the estimates obtained in Asencio {\it et. al.} (2014) in which truncation at 40 seconds was selected by cross validation, but where we believe the use of an explicit smoothing penalty may have biassed the results.


\bs

\cl{FIGURE 2}
\vskip 0.5cm
\moveright 0.5cm
\hbox{
\includegraphics[height=5cm]{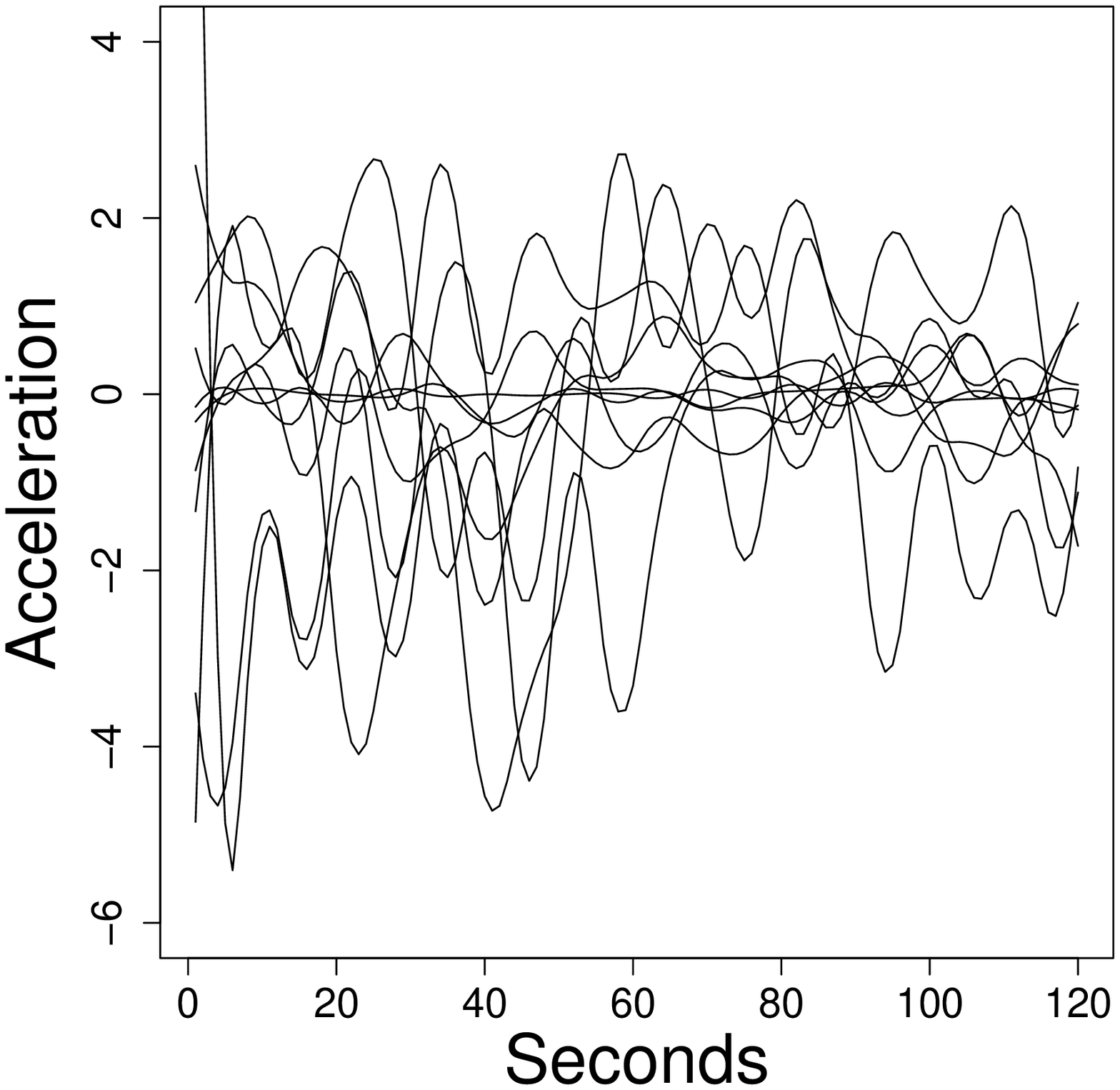} \hskip 0.5cm
\includegraphics[height=5cm]{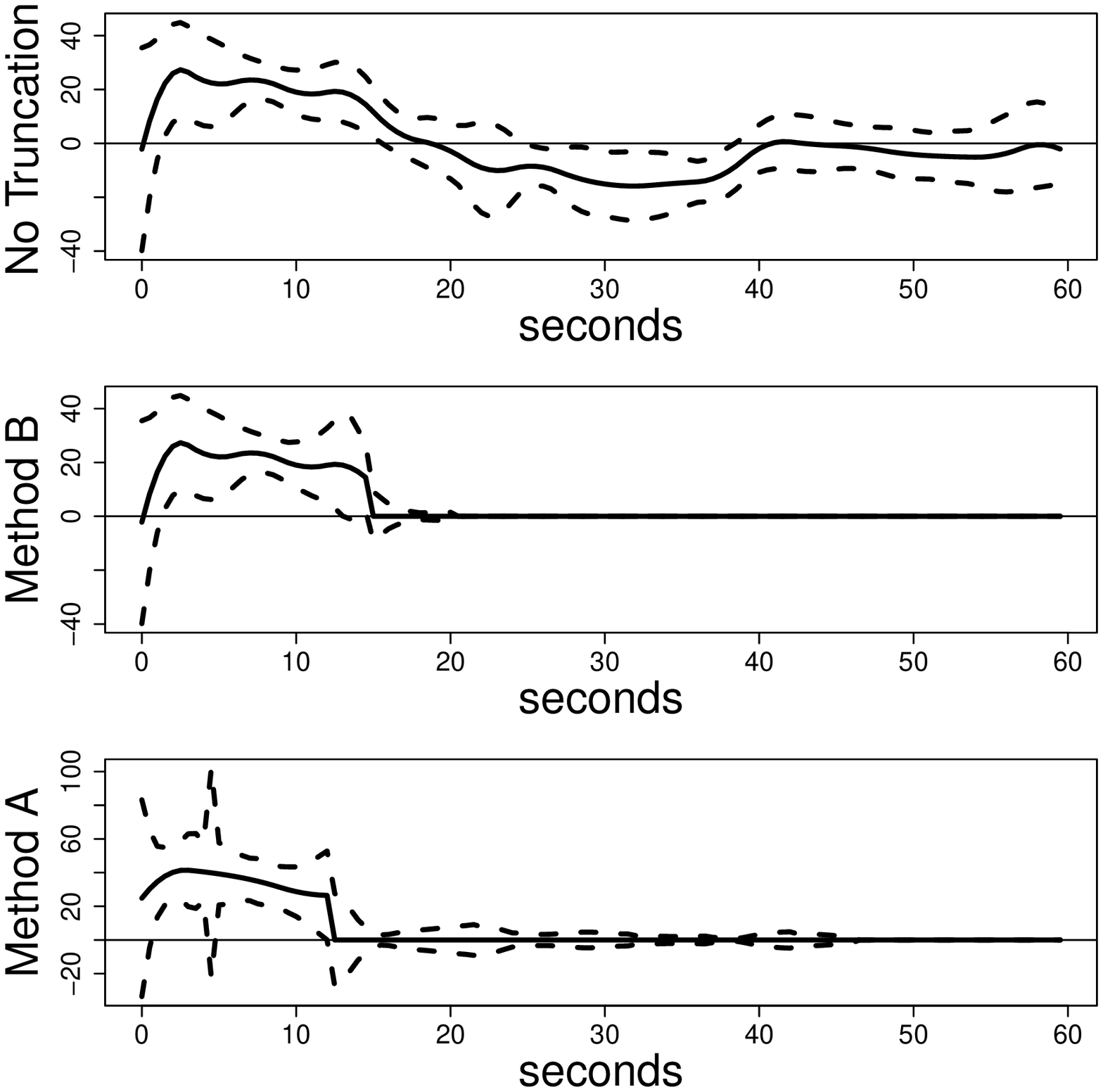}
}
{\leftskip = 0.5cm
\rightskip = 0.5cm
\ni Results of an analysis of PM data. Left: a sample of 10 covariate curves employed in the model. Right from top to bottom: estimates for $b$ with no truncation, using Method~B and using Method~A. Dashed lines give two-standard-deviation pointwise confidence intervals obtained from 200 residual bootstrap replicates based on the model from Method~A.
\par
}

\bs

\cl{\bf 6. THEORETICAL PROPERTIES}

\ni{\sl 6.1. Main result.}
In this section we show that the second method suggested in section~3.1 gives consistent estimators of $\th_0$, denoting the true value of~$\th$. Taking $\cI=[0,1]$, $u=0$ and $v=\th_0\in(0,1]$, we assume that the truncated linear model at (1.2) is correct. We write $a_0$ and $b_0$ for the true values of the scalar $a$ and function $b$, and assume of $b_0$ that:
$$
\eqalign{
&\hbox to4.42in{$b_0$ is continuous on $\cI$, vanishes on $[\th_0,1]$ and is zero at at most a finite}\cr
\noalign{\vskip-7pt}
&\hbox{number of points in $[0,\th_0]$.}\cr}\eqno(6.1)
$$
Taking $\th_1,\th_2$ to satisfy $0\leq\th_1<\th_0<\th_2\leq1$, we further assume that
$$
\sup_{\th_1\leq\th\leq\th_2}\,
\oon\,\sumion\bigg[\int_0^\th\big\{\cb(t)-b_0(t)\big\}\,\{X_i(t)-\bX(t)\}\,dt\bigg]^2
=O_p\big(\eta_n^2\big)\,,\eqno(6.2)
$$
where the positive, deterministic sequence $\eta_n$ satisfies $\eta_n=o(1)$ and $n\mo=O(\eta_n^2)$ as $n\rai$. The first of these conditions on $\eta_n$ merely reflects the consistency of $\cb$ for $b_0$, and the second is particularly mild since we would not expect $\eta_n$ to converge to zero faster than $n\mhf$.

On occasion we suppose in addition that
$$
\sup_{\th_1\leq\th\leq\th_2}\,
\bigg|\oon\,\sumion(\ep_i-\bep)\int_0^\th\big\{\cb(t)-b_0(t)\big\}\,\{X_i(t)-\bX(t)\}\,dt\bigg|
=O_p\big(n\mhf\,\eta_n\big)\,.\eqno(6.3)
$$
In the case of the standard methods discussed in section~3.2, arguments similar to those of Hall and Hosseini-Nasab (2009) can be used to prove that (6.2) and (6.3) hold for the same~$\eta_n$.

The linear model is assumed to be fitted by least-squares, and so the intercept, $a$, is estimated by $\ca=\bY-\inti\cb\,\bX$. In particular, $\ta$ is determined once we have computed the estimator $\tb$, and the expected value of $Y_i$, conditional on $X_i$, is modelled as $\bY+\inti\cb\,(X_i-\bX)$. Therefore it is not necessary to impose analogues of the constraints (6.2) and (6.3) on $\ca$ as well as on~$\cb$.

We suppose too that the errors $\ep_i$ are independent and identically distributed, and are independent of the $X_i$s; and that second moments are finite:
$$
\inti E\big(X^2\big)+E\big(\ep^2\big)<\infty\,,\quad E(\ep)=0\,.
\eqno(6.4)
$$
Then the singular-value decomposition at (2.7) is well defined, with eigenvalues and eigenvectors $\om_j$ and $\phi_j$, respectively. We assume that the eigenvalues, which form a nonincreasing sequence of positive numbers, decay sufficiently fast to ensure that
$$
\sumjoi\om_j\half\,\sup_{t\in\cI}\,|\phi_j(t)|<\infty\,.\eqno(6.5)
$$
Let $\hth$ denote any value of $\th$ that minimises $T(\th)$, at (3.5), on the interval $[\th_1,\th_2]$, where $\th_1$ and $\th_2$ are as in (6.2) and~(6.3).

\proclaim Theorem~6.1. (i)~If $(2.8)$, $(6.1)$, $(6.2)$, $(6.4)$ and $(6.5)$ hold, and if the penalty parameter $\la=\la(n)$ in $(3.5)$ satisfies $\la\ra0$ and $\la/\eta_n\rai$ as $n\rai$, then $\hth$ converges in probability to~$\th_0$ as $n\rai$. (ii)~If $(2.8)$ and $(6.1)$--$(6.5)$ obtain, and if the penalty parameter $\la$ in $(3.5)$ satisfies $\la\ra0$ and $\la/\eta_n^2\rai$ as $n\rai$, then the conclusion of part~(i) again holds.

\ni{\sl 6.2. Discussion.}
The reason for stating Theorem~6.1 in two versions, one when (6.3) holds and the other without imposing that constraint, is that the conditions on $\la$ are less stringent in the presence of~(6.3). In particular, $\la$ can decrease to zero more quickly when (6.3) holds, reflecting the fact that the assumption $\la/\eta_n\rai$ in part~(i) of the theorem is relaxed to $\la/\eta_n^2\rai$ in part~(ii).

Condition (6.2) typically holds for a choice of $\eta_n$ that decreases at a polynomial rate in $n\mo$ as $n\rai$. This reflects the fact that, in a large class of problems, prediction under the linear model can be undertaken with a polynomial level of accuracy, even if we do not have parametric models for the function $b_0$ or for the distributions of the functions $X_i$ or errors~$\ep_i$. See Cai and Hall (2006), particularly their Theorem~3.1.

The methods used by Cai and Hall (2006) can be employed to establish (6.2) for $\eta_n=n^{\eta-(1/2)}$, for any given $\eta\in(0,\thf)$, provided that, for example, the random functions $X$ have sufficiently many finite moments; the eigenvalues $\om_j$ in (2.7) decrease to zero in asymptotic proportion to $j^{-c}$, where $c$ is sufficiently large (depending on $\eta$); the spacings $\om_j-\om_{j+1}$ of the eigenvalues are no less than a fixed constant multiple of $j^{-(c+1)}$; the function $b_0$ admits a sufficiently rapidly convergent generalised Fourier expansion in terms of the eigenfunctions~$\phi_j$; and the orthonormal functions $\psi_j$ are taken to be the empirical versions, $\hphi_j$, of the~$\phi_j$s.

When using conventional methods to compute $\cb$, as outlined in section~3.2; and under the assumptions discussed in the previous paragraph; both (6.2) and (6.3) can be shown to hold for values $\eta_n$ that decrease at a polynomial rate, in particular at rate $n^{\de-(1/2)}$ where $\de\in(0,\thf)$. We outline details in the last paragraph of this section.

It is possible to establish an upper bound to the rate of convergence of $\hth$ to~$\th_0$. The upper bound, and also the actual convergence rate, decreases to zero more slowly, as $n\rai$, if we decrease the rate of convergence of $b_0(t)$ to zero as $t\uparrow\th_0$; that is, if $b_0$ becomes smoother at~$\th_0$. Here, smoothness can be characterised in terms of the number of bounded derivatives enjoyed by $b_0$ at~$\th_0$; the greater the number, the slower the convergence rate of $\hth$ to~$\th_0$.

To appreciate how (6.3) can be proved if $\cb$ is constructed as suggested in section~3.2, recall that $\cb=\sum_{k\leq m}\,\cbe_k\,\psi_k$, and note the formula for $\cbe_j$ at (3.7). In that notation, the quantity being bounded at (6.3) is given by
$$
\eqalignno{
&\sumion(\ep_i-\bep)\int_0^\th\big\{\cb(t)-b_0(t)\big\}\,\{X_i(t)-\bX(t)\}\,dt\cr
&=\sumion\ep_i\int_0^\th
\bigg\{\sum_{k=1}^m\,\bigg(B_k+\oon\,\sum_{i_1=1}^n\,
B_{i_1k}\,\ep_{i_1}\bigg)\,\psi_k(t)
-b_0(t)\bigg\}\,\{X_i(t)-\bX(t)\}\,dt\cr
&=\sumion\ep_i\int_0^\th
\bigg\{\sum_{k=1}^m\,B_k\,\psi_k(t)
-b_0(t)\bigg\}\,\{X_i(t)-\bX(t)\}\,dt\cr
&\qquad
+\oon\,\sumion\ep_i^2\int_0^\th
\bigg\{\sum_{k=1}^m\,B_{ik}\,\psi_k(t)\bigg\}\,\{X_i(t)-\bX(t)\}\,dt\cr
&\qquad
+\sumion\ep_i\int_0^\th
\bigg\{\bigg(\oon\,\sum_{i_1\,:\,i_1\neq i}\,
B_{i_1k}\,\ep_{i_1}\bigg)\,\psi_k(t)\bigg\}\,\{X_i(t)-\bX(t)\}\,dt\,,&(6.6)
}
$$
where all but the quantities that are written explicitly as $\ep_i$ or $\ep_{i_1}$ are measurable in the sigma-field generated by $\cX$, and so are conditioned upon and are independent of the errors $\ep_1,\ldots,\ep_n$. In consequence, the methods developed by Cai and Hall (2006) can be used to establish~(6.3).

\bs

\cl{\bf A. TECHNICAL ARGUMENTS}

\ni{\sl A.1. Proof of Theorem~2.1.}
Since, in view of (1.4), $D_2(h)=D_3(h)+E(\ep^2)$, where $D_3(h)=E\{g(X)-h(X)\}^2$, then $h$ is chosen equivalently to minimise $D_3(h)$; and since, by (2.2), $h$ can be translated to an arbitrary extent, then $E\{g(X)-h_0(X)\}=0$ at the minimum, i.e.~the first part of (2.5) must hold. Hence, without loss of generality, $E\{g(X)\}=0$. Therefore we wish to minimise
$$
D_3(h)=E\{g(X)-h(X)\}^2=E\big\{g(X)^2\big\}-2\,E\{g(X)\,h(X)\}+E\big\{h(X)^2\big\}\,,
$$
in the class $\cH_1\subseteq\cH$ of functions $h\in\cH$ such that $E\{h(X)\}=0$. We claim that any $h_0\in\cH_1$ that minimises $D_2(h_1)$ over that class satisfies the second part of~(2.5).

To appreciate why, suppose $h=h_0$ produces a minimum, and put $h_1=c\,h$ where $c$ is a constant. (Clearly, $h_1\in\cH_1$ whenever $h\in\cH_1$.) Then
$$
\eqalignno{
D_3(h_1)&=E\big\{g(X)^2\big\}-2\,c\,E\{g(X)\,h(X)\}+c^2\,E\big\{h(X)^2\big\}\,.
}
$$
Now, $D_3(h_1)$, treated as a function of $c$, is a convex parabola, and achieves a unique minimum when
$$
c=c_1\equiv{E\{g(X)\,h(X)\}\over E\{h(X)^2\}}\,.
$$
The second part of (2.5) must hold if $h$ there is replaced by~$h_2=c_1\,h$. This contradicts our assumption that $h$ produces a minimum, unless of course $c_1=1$, in which case the second part of (2.5) obtains. (That result is equivalent to $c_1=1$.)

\ni{\sl A.2. Proof of Theorem~2.2.}
Since any adjustment that centres $X$ can be accommodated in the scalars $a$ and $a_1$ in (2.6), then we can assume without loss of generality that $E(X)=0$. Then (2.6) implies that $a=a_1$ and
$$
E\bigg(\inti\de\,X\bigg)^{\!2}=0\,,\eqno(\A.1)
$$
where $\de$ denotes the function defined by $\de(t)=b(t)\,I(t\in[u,v])-b_1(t)\,I(t\in[u_1,v_1])$. Now, the left-hand side of (\A.1) is given by
$$
\inti\!\inti \de(t_1)\,\de(t_2)\,K(t_1,t_2)\,dt_1\,dt_2
=\sumjoi\om_j\,\bigg(\inti\de\,\phi_j\bigg)^{\!2}\,,
$$
where we used (2.7) to derive the identity. Therefore (\A.1) holds if and only if, for each $j$, $\om_j\half\inti\de\,\phi_j=0$, and in view of (2.8) this is equivalent to $\de=0$ almost everywhere on~$\cI$.

\ni{\sl A.3. Proof of Theorem~6.1.}

\ni{\sl Step~1: Preparatory lemma.}
Let $\th_1,\th_2$ be as in (6.2) and~(6.3). That is, $0\leq\th_1<\th_0<\th_2\leq1$.

\proclaim Lemma. If $(6.1)$, $(6.4)$ and $(6.5)$ hold then, uniformly in $\th\in[\th_1,\th_2]$,
$$
\sumion(\ep_i-\bep)\int_\th^{\th_0}b_0(t)\,\{X_i(t)-\bX(t)\}\,dt
=O_p\big(n\half\,|\th-\th_0|\big)\,.\eqno(\A.2)
$$

\ni{\sl Proof.}
We show that, uniformly in $\th\in[\th_1,\th_2]$,
$$
\sumion\ep_i\int_\th^{\th_0}b_0(t)\,\{X_i(t)-\mu(t)\}\,dt
=O_p\big(n\half\,|\th-\th_0|\big)\,,\eqno(\A.3)
$$
where $\mu=E(X)$. A similar but simpler argument demonstrates that, uniformly in the same sense,
$$
\bep\,\sumion\int_\th^{\th_0}b_0(t)\,\{X_i(t)-\mu(t)\}\,dt
=O_p\big(n\half\,|\th-\th_0|\big)\,,\eqno(\A.4)
$$
and together, (\A.3) and (\A.4) imply that
$$
\sumion(\ep_i-\bep)\int_\th^{\th_0}b_0(t)\,\{X_i(t)-\mu(t)\}\,dt
=O_p\big(n\half\,|\th-\th_0|\big)\,,
$$
which, since $\sumi(\ep_i-\bep)\,\int_\th^{\th_0}b_0\,(\bX-\mu)=0$, is equivalent to (\A.2).

To derive (\A.3), first recall the singular-value decomposition at (2.7), involving the (eigenvalue, eigenfunction) pairs $(\om_j,\phi_j)$. The same decomposition applies to the random function $\ep\,(X-\mu)$, except that the eigenvalues are now $\si^2\,\om_j$, where $\si^2=E(\ep^2)$. (The eigenfunctions are unchanged.) Therefore we can write
$$
\ep\,\{X(t)-\mu(t)\}=\sumjoi\xi_j\,\phi_j(t)\,,
$$
where the random variables $\xi_j$ have zero means and respective variances $\si^2\,\om_j$. Hence, writing $\bxi_j$ for the mean of $n$ random variables all independent and identically distributed as $\xi_j$, we have:
$$
\eqalignno{
\bigg|\oon\,\sumion\ep_i\int_\th^{\th_0}b_0(t)\,\{X_i(t)&-\mu(t)\}\,dt\bigg|
=\bigg|\int_\th^{\th_0}b_0(t)\,\bigg\{\sumjoi\bxi_j\,\phi_j(t)\bigg\}\,dt\bigg|\cr
&\leq|\th-\th_0|\,\bigg\{\sup_{t\in\cI}\,|b_0(t)|\bigg\}\,\sumjoi|\bxi_j|\,
\sup_{t\in\cI}\,|\phi_j(t)|\,.\qquad&(\A.5)\cr
}
$$
Writing $s_j=\sup_{t\in\cI}\,|\phi_j(t)|$, we have:
$$
E\bigg(\sumjoi|\bxi_j|\,s_j\bigg)
\leq\sumjoi\big(E\bxi_j^2\big)\half\,s_j
=n\mhf\,\si\,\sumjoi\om_j\half\,s_j
=O\big(n\mhf\big)\,,\eqno(\A.6)
$$
where the last identity follows from~(6.5). Result (\A.3) is a consequence of (\A.5) and~(\A.6). Note that the fact that $b_0$ is continuous on $\cI$, as assumed in (6.1), implies that $\sup_{t\in\cI}\,|b_0(t)|<\infty$.

\ni{\sl Step~2: Expansions of $T(\th)$.} The two expansions, first in cases where both (2.7) and (6.3) hold, and secondly where only (6.2) obtains, are given at (\A.9) and (\A.11), respectively.

Since $\cI=[0,1]$ and $Y_i=E(Y)+\int _{[0,\th_0]}b_0\,(X_i-EX)+\ep_i$ then $Y_i-\bY=\int_{[0,\th_0]}b_0\,(X_i-\bX)+\ep_i-\bep$, and hence,
$$
\eqalignno{
T(\th)&-n\,\la\,\th=\sumion\bigg\{Y_i-\ca-\int_0^\th\cb(t)\,X_i(t)\,dt\bigg\}^2\cr
&=\sumion\bigg[Y_i-\bY-\int_0^\th\cb(t)\,\{X_i(t)-\bX(t)\}\,dt\bigg]^2\cr
&=\sumion\bigg[\int_0^{\th_0}b_0(t)\,\{X_i(t)-\bX(t)\}\,dt
-\int_0^\th\cb(t)\,\{X_i(t)-\bX(t)\}\,dt+\ep_i-\bep\bigg]^2\cr
&=\sumion\bigg[\int_\th^{\th_0}b_0(t)\,\{X_i(t)-\bX(t)\}\,dt\cr
&\qquad\qquad
-\int_0^\th\big\{\cb(t)-b_0(t)\big\}\,\{X_i(t)-\bX(t)\}\,dt+\ep_i-\bep\bigg]^2\cr
&=\sumion\Bigg(\bigg[\int_\th^{\th_0}b_0(t)\,\{X_i(t)-\bX(t)\}\,dt\bigg]^2\cr
&\qquad\qquad
+\bigg[\int_0^\th\big\{\cb(t)-b_0(t)\big\}\,\{X_i(t)-\bX(t)\}\,dt\bigg]^2\Bigg)\cr
&\qquad\qquad
+2\,\sumion\bigg[\int_\th^{\th_0}b_0(t)\,\{X_i(t)-\bX(t)\}\,dt\cr
&\qquad\qquad\qquad
-\int_0^\th\big\{\cb(t)-b_0(t)\big\}\,\{X_i(t)-\bX(t)\}\,dt\bigg]\,(\ep_i-\bep)
+\sumion(\ep_i-\bep)^2\cr
&=T_4(\th)+\sumion(\ep_i-\bep)^2
+O_p\Big[n\half\,\big\{|\th-\th_0|\,I(\th)+\eta_n\big\}+n\eta_n^2\Big]\,,&(\A.7)
}
$$
uniformly in $\th\in[\th_1,\th_2]$, where $I(\th)=I(\th<\th_0)$,
$$
T_4(\th)=n\int_\th^{\th_0}\!\!\int_\th^{\th_0}b_0(t_1)\,b_0(t_2)\,\hK(t_1,t_2)\,dt_1\,dt_2\,,
\eqno(\A.8)
$$
and $\hK$ is as at~(3.2). To obtain the last identity in (\A.7) we used formulae (6.2), (6.3) and~(\A.2).

Using the fact that, as assumed immediately below (T.2), $n\mo=O(\eta_n^2)$, we deduce that $n\half\eta_n=O(n\eta_n^2)$. Therefore (\A.7) entails:
$$
n\mo\,T(\th)=n\mo\,T_4(\th)+\oon\,\sumion(\ep_i-\bep)^2+\la\,\th^2
+O_p\big\{n\mhf\,|\th-\th_0|\,I(\th)+\eta_n^2\big\}\,.\eqno(\A.9)
$$

Of course, (6.3) is assumed only in part~(ii) of Theorem~6.1. In the statement of part~(i) of the theorem we impose condition (6.2) but not (6.3), but from (6.2) we can derive the following bounds, uniformly in $\th\in[\th_1,\th_2]$, for the quantity of which the absolute value is taken on the left-hand side of~(6.3):
$$
\eqalignno{
\bigg|\sumion(\ep_i&-\bep)\int_0^\th\big\{\cb(t)-b_0(t)\big\}\,\{X_i(t)-\bX(t)\}\,dt\bigg|\cr
&\leq\Bigg(\sumion(\ep_i-\bep)^2\,\cdot\,
\sumion\bigg[\int_0^\th\big\{\cb(t)-b_0(t)\big\}\,\{X_i(t)-\bX(t)\}\,dt\bigg]^2\Bigg)^{\!1/2}\cr
&=O_p\big\{(n\cdot n\eta_n^2)\half\big\}
=O_p\big(n\eta_n\big)\,.&(\A.10)
}
$$
Using (\A.10) in place of (6.3), but in all other respects using the argument leading to (\A.7), we obtain the following expansion in place of~(\A.9):
$$
\eqalignno{
n\mo\,T(\th)
=n\mo\,T_4(\th)&+\oon\,\sumion(\ep_i-\bep)^2+\la\,\th^2\cr
&+O_p\Big\{n\mhf\,|\th-\th_0|\,I(\th)+\eta_n\Big\}\,,&(\A.11)
}
$$
uniformly in $\th\in[\th_1,\th_2]$.

\ni{\sl Step~3: Completion.}
Combining (\A.9) and (\A.11) we deduce that
$$
\eqalignno{
n\mo\,T(\th)
=n\mo\,T_4(\th)&+n\mo\,\sumion(\ep_i-\bep)^2+\la\,\th^2
+O_p\big\{n\mhf\,|\th-\th_0|\,I(\th)\big\}\cr
&+\cases{O_p(\eta_n^2)&if (6.2) and (6.3) hold\cr
O_p(\eta_n)&if only (6.2) holds$\,,$}
\quad&(\A.12)
}
$$
where the remainders are of the stated sizes uniformly in $\th\in[\th_1,\th_2]$. If $\th>\th_0$ then, in view of (6.1) and the definition of $T_4(\th)$ at (\A.8), $T_4(\th)=0$, and so (\A.12) simplifies to:
$$
n\mo\,T(\th)=U+\la\,(\th^2-\th_0^2)
+\cases{O_p(\eta_n^2)&if (6.2) and (6.3) hold\cr
O_p(\eta_n)&if only (6.2) holds$\,,$}
$$
where the random variable $U$ does not depend on $\th$, and now the remainders are of the stated sizes uniformly in $\th\in[\th_0,\th_2]$. Hence, since $\la/\eta_n^2\rai$ if (6.2) and (6.3) both hold (i.e.~if we are in the context of part~(i) of Theorem~6.1); and since $\la/\eta_n\rai$ if only (6.2) is assumed (i.e.~if we are in the context of part~(i) of the theorem); and if $\hth$ is chosen to minimise $T(\th)$ for $\th\in[\th_1,\th_2]$; then, for each $\de>0$, $P(\hth>\th_0+\de)\ra0$.

On the other hand, if $\th<\th_0$ then (\A.12) implies that
$$
\eqalignno{
n\mo\,T(\th)&=n\mo\,T_4(\th)+V-\la\,(\th_0^2-\th^2)\cr
&\qquad
+\cases{O_p\{n\mhf\,(\th_0-\th)+\eta_n^2\}&if (6.2) and (6.3) hold\cr
O_p\{n\mhf\,(\th_0-\th)+\eta_n\}&if only (6.2) holds$\,,$}\qquad&(\A.13)
}
$$
where the random variable $V$ does not depend on $\th$. Define
$$
\ka(\th)=\int_\th^{\th_0}\!\int_\th^{\th_0}b_0(t_1)\,b_0(t_2)\,
K(t_1,t_2)\,dt_1\,dt_2\,.
$$
In the next paragraph we show that, for any bounded function $f$ on~$\cI$,
$$
\sup_{u,v\in\cI}\,\bigg|\int_u^v\!\int_u^v
f(t_1)\,f(t_2)\,\big\{\hK(t_1,t_2)-K(t_1,t_2)\big\}\,dt_1\,dt_2\bigg|
\ra0\,,\eqno(\A.14)
$$
where the convergence is in probability. Taking $f=b_0$ we deduce from the definition of $T_4(\th)$ at (\A.8) that
$$
\sup_{\th\,:\,\th\in[\th_1,\th_0]}\,\big|n\mo\,T_4(\th)-\ka(\th)\big|\ra0
\eqno(\A.15)
$$
in probability. It follows from (6.1) that $\ka(\th)$ is strictly positive whenever $\th<\th_0$. Using this property, (\A.13), (\A.15) and the fact that $\la\ra0$ as $n\rai$, we deduce that for each $\de>0$, $P(\hth<\th_0-\de)\ra0$. Combining this result with the property that $P(\hth>\th_0+\de)\ra0$, derived in the previous paragraph, we deduce that $\hth\ra\th_0$ in probability, as had to be proved.

To derive (\A.14), let $\lhs$ denote the left-hand side of (\A.14), and note that, since $\cI=[0,1]$, then $\lhs\leq(\sup|f|)^2\,\hJ$, where
$$
\hJ\equiv\inti\!\inti\big|\hK(t_1,t_2)-K(t_1,t_2)\big|\,dt_1\,dt_2\,.
$$
Since $\inti E(X^2)<\infty$ (see (6.4)) then $E\{|\hK(t_1,t_2)-K(t_1,t_2)|\}\ra0$ as $n\rai$, for each pair $t_1,t_2\in\cI$. Similarly, $E(\hJ)\ra0$ as $n\rai$. Hence, $\hJ\ra0$ in probability, implying that $\lhs\ra0$ in probability, i.e.~(\A.14) holds.

\bs

\cl{\bf REFERENCES}
\frenchspacing

\baselineskip=13pt

\nh APANASOVICH, T.V. AND GOLDSTEIN, E. (2008). On prediction error in functional linear regression. {\sl Statist. Probab. Lett.} {\bf 78}, 1807--1810.

\nh ASENCIO, M., HOOKER, G. AND GAO H.O., (2014), Functional Convolution Models. {\sl Stat. Mod.} {\bf 14}, 1-21.

\nh BA\'ILLO, A. (2009). A note on functional linear regression. {\sl J. Stat. Comput. Simul.} {\bf 79}, 657--669.

\nh CAI, T.T. AND HALL, P. (2006). Prediction in functional linear regression. {\sl Ann. Statist.} {\bf 34}, 2159--2179.

\nh CAI, T.T. AND YUAN, M. (2012). Minimax and adaptive prediction for functional linear regression. {\sl J. Amer. Statist. Assoc.} {\bf 107}, 1201--1216.

\nh CAI, T.T. AND ZHOU, H.H. (2013). Adaptive functional linear regression. {\tt http:\break //www-stat.wharton.upenn.edu/$\sim$tcai/paper/Adaptive-FLR.pdf}.

\nh CARDOT, H., FERRATY, F. AND SARDA, P. (1999). Functional linear model. {\sl Statist. Probab. Lett.} {\bf 45}, 11--22.

\nh CARDOT, H., FERRATY, F. AND SARDA, P. (2003). Spline estimators for the functional linear model. {\sl Statist. Sinica} {\bf 13}, 571--591.

\nh CARDOT, H., MAS, A. AND SARDA, P. (2007). CLT in functional linear regression models. {\sl Probab. Theory Related Fields} {\bf 138}, 325--361.

\nh CLARK, N. N., GAUTAM, M.,  WAYNE, W. S., LYONS, D. W., THOMPSON, G. AND ZIELINSKA, B. (2007).
Heavy-duty chassis dynamometer testing for emissions inventory, air quality modeling, source
apportionment and air toxins emissions inventory: E55/59 all phases. {\sl Technical Report E55/59,
Coordinating Research Council.}

\nh COMTE, F. AND JOHANNES, J. (2012). Adaptive functional linear regression. {\sl Ann. Statist.} {\bf 40}, 2765--2797.

\nh CRAMBES, C., KNEIP, A. AND SARDA, P. (2008). Estimation of the functional linear regression with smoothing splines. In {\sl Functional and Operatorial Statistics}, eds S. Dabo-Niang and F. Ferraty, pp.~117--120. {\sl Contrib. Statist.}, Physica-Verlag/Springer, Heidelberg.

\nh CRAMBES, C., KNEIP, A. AND SARDA, P. (2009). Smoothing splines estimators for functional linear regression. {\sl Ann. Statist.} {\bf 37}, 35--72.

\nh FAN, J. AND ZHANG, J.-T. (2000). Two-step estimation of functional linear models with applications to longitudinal data. {\sl J. R. Stat. Soc.} Ser.~B {\bf 62}, 303--322.

\nh FERRATY, F., GONZ\'ALEZ-MANTEIGA, W., MART\'INEZ-CALVO, A. AND\break VIEU, P. (2012). Presmoothing in functional linear regression. {\sl Statist. Sin\-ica} {\bf 22}, 69--94.

\nh HALL, P. AND HOROWITZ, J.L. (2007). Methodology and convergence rates for functional linear regression. {\sl Ann. Statist.} {\bf 35}, 70--91.

\nh HALL, P. AND HOSSEINI-NASAB, M. (2006). On properties of functional principal components analysis. {\sl J. R. Stat. Soc.} Ser.~B {\bf 68}, 109--126.

\nh HALL, P. AND HOSSEINI-NASAB, M. (2009). Theory for high-order bounds in functional principal components analysis. {\sl Math. Proc. Cambridge Philos. Soc.} {\bf 146}, 225--256.

\nh HE, G., M\"ULLER, H.-G. AND WANG, J.-L. (2010). Functional linear regression via canonical analysis. {\sl Bernoulli} {\bf 16}, 705--729.

\nh JAMES, G., WANG, J. AND ZHU, J. (2009). Functional linear regression that's interpretable. {\sl Ann. Statist.} {\bf 37}, 2083--2108.

\nh JOHANNES, J. AND SCHENK, R. (2012). Adaptive estimation of linear functionals in functional linear models. {\sl Math. Methods Statist.} {\bf 21}, 189--214.

\nh JOHANNES, J. AND SCHENK, R. (2013). On rate optimal local estimation in functional linear model. {\sl Electronic J. Statist.} {\bf 7}, 191--216.

\nh LI, Y. AND HSING, T. (2007). On rates of convergence in functional linear regression. {\sl J. Multivariate Anal.} {\bf 98}, 1782--1804.

\nh MARONNA, R.A. AND YOHAI, V.J. (2013). Robust functional linear regression based on splines. {\sl Comput. Statist. Data Anal.} {\bf 65}, 46--55.

\nh MAS, A. AND PUMO, B. (2009). Functional linear regression with derivatives. {\sl J. Nonparametr. Statist.} {\bf 21}, 19--40.

\nh MCLEAN, M. W., HOOKER, G. AND RUPPERT, D., (2014). Restricted Likelihood Ratio Tests for Linearity in Scalar-on-Function Regression. {\sl Stat. Comp.} in press.

\nh RAMSAY, J.O. AND SILVERMAN, B.W. (2002). {\sl Applied Functional Data Analysis}. Springer, New York.

\nh RAMSAY, J.O. AND SILVERMAN, B.W. (2005). {\sl Functional Data Analysis}, Second Edn. Springer, New York.

\nh YAO, F., M\"ULLER, H.-G. AND WANG, J.-L. (2005). Functional linear regression analysis for longitudinal data. {\sl Ann. Statist.} {\bf 33}, 2873--2903.

\nh YUAN, M. AND CAI, T.T. (2010). A reproducing kernel Hilbert space approach to functional linear regression. {\sl Ann. Statist.} {\bf 38}, 3412--3444.

\nh WU, Y., FAN, J. AND M\"ULLER, H.-J. (2010). Varying-coefficient functional linear regression. {\sl Bernoulli} {\bf 16}, 730--758.

\vend